\documentclass[12pt,english,reviewcopy]{elsart}
\usepackage[T1]{fontenc}
\usepackage[latin1]{inputenc}
\usepackage{rotating}
\usepackage{color}
\usepackage{graphicx}

\makeatletter

\usepackage{babel}
\makeatother
\journal{Nuclear Physics A}
\begin{document}

\begin{frontmatter}
\title{A Genetic Algorithm Analysis of $N^{*}$ Resonances in
$p\left(\gamma,K^{+}\right)\Lambda$ Reactions }

\author{D.G.~Ireland\corauthref{cor}},
\corauth[cor]{Corresponding author.}
\ead{d.ireland@physics.gla.ac.uk} 
\address{Department of Physics and Astronomy, University of
Glasgow, Glasgow, G12 8QQ, Scotland, UK}

\author{S.~Janssen},
\author{J.~Ryckebusch}
\address{Department of Subatomic and Radiation Physics, Ghent
University, Proeftuinstraat 86, B-9000 Gent, Belgium}

\begin{abstract}
The problem of extracting information on new and known $N^{*}$
resonances by fitting isobar models to photonuclear data is
addressed. A new fitting strategy, incorporating a genetic algorithm,
is outlined. As an example, the method is applied to a typical
tree-level analysis of published $p\left(\gamma,K^{+}\right)\Lambda$
data. It is shown that, within the limitations of this tree-level
analysis, a resonance in addition to the known set is required to
obtain a reasonable fit. An additional $P_{11}$ resonance, with a mass
of about 1.9 GeV, gives the best agreement with the published data,
but additional $S_{11}$ or $D_{13}$ resonances cannot be ruled
out. Our genetic algorithm method predicts that photon beam asymmetry
and double polarization $p\left(\gamma,K^{+}\right)\Lambda$
measurements should provide the most sensitive information with
respect to missing resonances.
\end{abstract}


\end{frontmatter}

\section{Introduction}

The spectroscopy of baryons continues to be a subject of great interest
in intermediate energy nuclear physics, since it an essential component
in underpinning our knowledge of the substructure of the nucleon.
Central to this is the issue of which resonances exist. Constituent
quark models such as those proposed by Capstick and Roberts \cite{Capstick}
predict a large number of resonances which thus far have not been
observed, whilst for models which limit quark degrees of freedom \cite{oet00},
the number of {}``missing'' resonances is less.

The majority of resonance information has been gleaned from analyses
of single pion production reactions \cite{PDG}, and the suggestion
that the missing resonances may couple more strongly to other channels
\cite{Capstick,Capstick_2} has led to a number of experiments being
proposed and carried out at intermediate energy accelerator facilities.
Of particular interest is the possibility of studying strangeness
production, since this opens up the possibility of extracting extra
$N{}^{*}$ information.

The use of effective field theories is necessary in the resonance
region since QCD cannot be solved perturbatively at this energy scale.
Constituent quark models are able to predict quantities such as coupling
constants and electromagnetic form factors, but it is not straightforward
for these models to predict real observables, as a fundamental understanding
of the underlying dynamics is missing. By using an effective field
theory to extract coupling constants from fits to data, one can obtain
numbers to compare with quark model predictions. 

A complete understanding of the physics underlying the photonucleon
data in the resonance region will in principle only be possible in
a framework that contains all participating processes which include
coupling to resonances \cite{Manley,Vrana,Penner,chi01,kai97}. This means
taking into account meson production reactions such as $\gamma
N\rightarrow\pi N,\:\pi\pi N,\:\eta N,\:\omega N,\: K\Lambda,\:
K\Sigma,...,$ 
as well as Compton scattering and meson-induced production reactions.
This is clearly an enormous task. 

In a model adopting effective degrees-of-freedom, the field associated
with each resonance has a number of free parameters which are usually
determined by fitting model calculations to data. In a full-blown
coupled-channels approach, the number of free parameters could easily
end up being over 100. For such a procedure to be possible, a sizeable
data set for each of the contributing channels is required. One
difficulty, which is often overlooked, is the process of how to obtain
a set of parameters which leads to the best description of the
data. Related to this is the issue of how to assign error bars and
confidence levels to the values of the extracted resonance parameters.

An analysis of a single channel, whilst not complete, offers the
possibility of simplifying the problem by reducing the number of free
parameters to a manageable size, and in being able to identify the
most important features. Recent attempts to analyse the
$p(\gamma,K^{+})\Lambda$ channel in such a framework have highlighted
the tantalising prospect that previously undiscovered resonances may
reveal themselves in the mechanisms involving strangeness
production. Mart and Bennhold \cite{Mart2} used their tree-level
hadrodynamical model to show that the total cross-section measured at
SAPHIR \cite{Tran} could be reproduced by introducing a $D_{13}(1895)$
resonance. This resonance does not appear in the Particle Data Group
baryon summary table as it had not been conclusively observed, but was
included in \cite{Mart2} because of the predicted coupling
\cite{Capstick} to strange channels. In addition to this, a recent
re-analysis of pion photoproduction data \cite{cra02} appeared to find
evidence for it. On the other hand, an analysis by Saghai
\cite{Saghai_hyp_res} argued that by tuning the background processes
involved, the need for the extra resonance was removed. Both
approaches were based on the analysis of the SAPHIR data set, which at
that point amounted to around 100 data points.

In a previous article \cite{jan03}, we highlighted the problem of
extracting reliable resonance information from the limited set of
$p(\gamma,K^{+})\Lambda$ data points by performing many independent
fitting calculations. Using a fitting strategy based on a genetic
algorithm we illustrated that a large number of different solutions
resulted in similar $\chi^{2}$ values. We were able to group these
solutions into two main sets: one with $D_{13}$ coupling constants
close to zero (indicating its non-existence), and one with significantly
non-zero $D_{13}$ coupling constants (indicating its existence).
We further showed that the measurement of polarization observables,
such as photon beam asymmetry, could have a pivotal role in removing
ambiguities.

We have now performed a systematic study into the feasibility of determining
the combination of resonances which contribute to the $p(\gamma,K^{+})\Lambda$
reaction. To this end, we have used a typical tree-level description
of the reaction process in combination with a minimization procedure
which is specifically designed to quantify the relative success of
different combinations of resonances. This procedure employs a genetic
algorithm which is able to search efficiently a large parameter space,
coupled with a conventional minimization routine to ensure convergence.
We have benefitted from adding new photon beam asymmetry measurements
from SPring-8 \cite{zeg03}, as well as fitting electroproduction
data measured at Jefferson Lab \cite{moh03}.

The structure of this paper is as follows. We outline the features
of a typical tree-level $p(\gamma,K^{+})\Lambda$ reaction model in
section \ref{sec:Model-Formalism}. Section \ref{sec:Analysis-Procedure}
describes a framework which enables the extraction of resonance information
from existing photonucleon data, and discusses how to evaluate the
reliability of the information. As an example, the methodology is
applied to a tree-level analysis of $p(\gamma,K^{+})\Lambda$ in section
\ref{sec:Results-and-discussion}. We discuss the results of our work
and highlight the important next steps in studying this problem. We
present conclusions in section \ref{sec:Conclusion}.

\section{\label{sec:Model-Formalism}Model Formalism}

\begin{figure}
\begin{center}
\includegraphics[%
  scale=0.7]{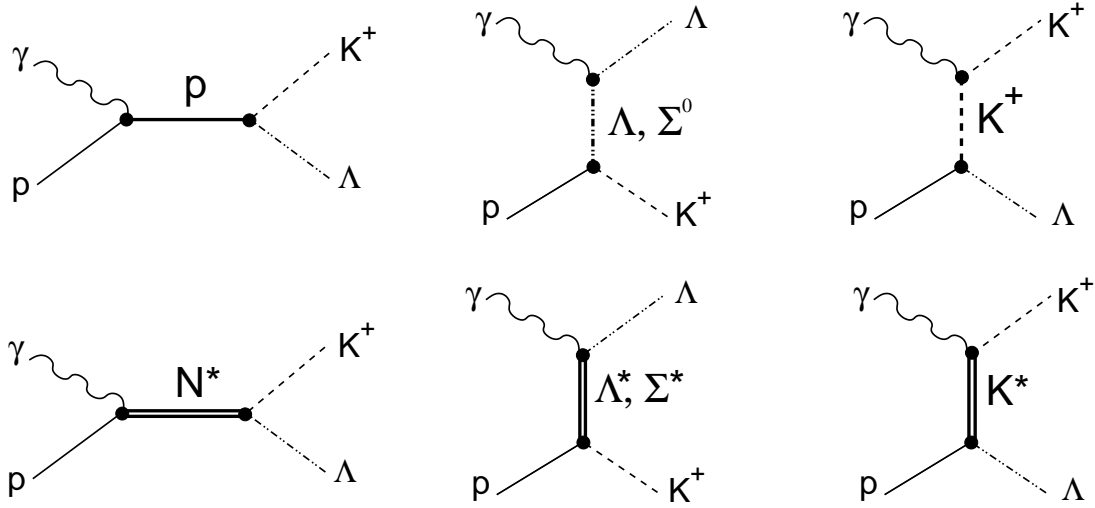}

\caption{\label{cap:Feynman-diagrams}Diagrams contributing to the
$p(\gamma,K^{+})\Lambda$ process at the tree level. The upper row
corresponds to the Born terms in which a proton is exchanged in the
\emph{s}-channel, a $\Lambda$ or $\Sigma^{+}$ in the \emph{u}-channel,
and a $K^{+}$ in the \emph{t}-channel. The lower row shows the
corresponding diagrams with the exchange of an excited particle or
resonance.}
\end{center}
\end{figure}

We adopt a hadrodynamical framework for modelling kaon photoproduction
on the nucleon. The model we use has been described in previous work
\cite{Janssen_role_hyp}, so we present only a brief summary here.
The $p(\gamma,K^{+})\Lambda$ reaction process is described by hadronic
degrees of freedom using an effective Lagrangian. The contributioning
diagrams are depicted in figure \ref{cap:Feynman-diagrams}. Every
intermediate particle in the reaction is treated as an effective field
with associated mass, photocoupling amplitudes and strong decay
widths. Tree level Feynman diagrams contain the vector $K^{*}(892)$
and the axial-vector $K_{1}(1270)$ \emph{t-}channel mesons, as well as
the usual Born terms. Two hyperon resonances, the $S_{01}(1800)$ and
$P_{01}(1810)$ are present in the \emph{u}-channel.  These ingredients
constitute the so-called {}``background''.

In Ref. \cite{Janssen_backgr} we stressed the difficulties associated
with parameterizing the background diagrams in $p(\gamma,K^{+})\Lambda$
calculations and presented results for three plausible background
schemes. Subsequent work \cite{Janssen_elec} showed that the $p(e,e'K^{+})\Lambda$
process is highly selective with respect to viable choices for dealing
with the background diagrams. The model used here is the only one
which we found to reproduce simultaneously the $p(\gamma,K^{+})\Lambda$
and $p(e,e'K^{+})\Lambda$ data. 

The finite extension of the meson-baryon vertices is implemented by
the use of hadronic form factors \cite{Haberzettl_gauge,Davidson,kor03}.
We also note in passing that the validity of this approach is only
likely to be reasonable for photon energies below about 2 GeV, but
its precise range of applicability remains to be established.

For the purposes of the present work, we have defined several versions
of our hadrodynamical model which correspond to different choices
of \emph{s}-channel resonances. For brevity, we subsequently refer
to these as different model variants, but the fact that they are based
on the same hadrodynamical framework should be kept in mind. Each
variant includes a {}``core'' set of resonances, consisting of $S_{11}(1650)$,
$P_{11}(1710)$ and $P_{13}(1720)$ which are well known to contribute
to the $p(\gamma,K^{+})\Lambda$ reaction \cite{Mart2,Saghai_hyp_res,Feuster2}.
The variant containing only these resonances is referred to as the
{}``Core'' variant. 

The other model variants each contain one other resonance of mass
1895 MeV. A $D_{13}$ resonance was chosen to be compatible with our
earlier work and because it had been proposed in \cite{Mart2}. This
in turn was motivated by the quark model of Capstick and Roberts \cite{Capstick}
which showed that the $D_{13}$ had the largest coupling strength
to the lowest spin states. However, rather than relying on the predictions
of a constituent quark model calculation, we have tried different
variants which contain a resonance of the same mass but alternative
quantum numbers. If another resonance does contribute, the exact value
of its mass is not likely to be important. Each of these variants
is thus referred to by the character of the additional resonance:
$S_{11}$, $P_{11}$, $P_{13}$ and $D_{13}$ . Our previous analysis
only used Core plus $D_{13}$ resonances \cite{jan03}.

We therefore have set up our programme of calculations to address
the questions of whether an additional resonance is required in the
description of the reaction (Core versus non-Core variants), and what
character an additional resonance might have ($S_{11}$, $P_{11}$,
$P_{13}$ and $D_{13}$ variants).

\section{\label{sec:Analysis-Procedure}Analysis Procedure}

For each model variant, we performed a set of calculations to extract
$N^{*}$ information from the reaction data. Each set consisted of
100 separate minimization calculations. The data sets employed in
the fitting procedure included total cross-sections, differential
photo-production cross-sections and recoil polarizations from the
SAPHIR data set \cite{Tran}, and photon beam-polarization asymmetries
from SPring-8 \cite{zeg03}. In addition, we also used separated longitudinal
and transverse electroproduction data from Jefferson Lab \cite{moh03},
as well as total electroproduction cross sections from a variety of
older sources \cite{Bebek_74,Bebek_77,Brown_cea}.

The coupling constants associated with the hadronic vertices in all
the amplitudes included in the model are free parameters in the
fitting procedure. A description of their exact definition is given in
\cite{stijn_thesis}, but we briefly list them here for completeness:
coupling constants related to Born terms, $g_{K^{+} \Lambda p}$ and
$g_{K^{+} \Sigma p}$; vector and axial vector mesons in the
$t-$channel, $G_{K^{*}}^{v}$, $G_{K^{*}}^{t}$, $G_{K_{1}}^{v}$ and
$G_{K_{1}}^{t}$; spin-$\frac{1}{2}$ $Y^{*}$ resonances in the
$u-$channel, $G_{Y^{*}Kp}$; spin-$\frac{1}{2}$ $N^{*}$ resonances in
the $s-$channel, $G_{N^{*}K \Lambda}$; spin-$\frac{3}{2}$ $N^{*}$
resonances in the $s-$channel, $G^{1}_{N^{*}K \Lambda}$,
$G^{2}_{N^{*}K \Lambda}$ and three off-shell parameters. Note that
these constants are products of both a hadronic part and an
electromagnetic part. Two hadronic cut-off parameters (one for all the
Born terms, $\Lambda_{born}$, and and for all resonances in the $s-$,
$t-$ and $u-$ channels, $\Lambda_{res}$) are also free parameters.

To compare the relative success of each model variant in describing
the data, we require two separate but related procedures: a search
for the best set of free parameters for each model variant, and a
comparison amongst the different model variants. The first procedure
is achieved by varying the free parameters and searching for a minimum
$\chi^{2}$ statistic by fitting calculations to the data. This is
a fairly standard procedure, but we remind the reader in passing the
assumptions which are implicitly required: the data points being fitted
must be independent (the value of one does not affect the others),
and error bars on the points represent the standard deviations of
a Gaussian probability density. Both these requirements are likely
to be approximately correct, and thus minimising $\chi^{2}$ is equivalent
to maximising a likelihood function. The likelihood represents the
probability that the data would be measured, given a particular set
of free parameters. 

Furthermore, there is the assumption that no particular values of
the free parameters are to be favoured prior to the fitting procedure.
By imposing limits on the free parameters (as we do through physics
constraints) this is not strictly true, but will be approximately
true if the limits are sufficiently large. The whole procedure is
then equivalent to maximising the probability that a particular set
of free parameters is correct, given the experimental data, by varying
the free parameters.

Comparing different model variants needs to be done carefully, since
each one may have a different number of free parameters. The approach
we take implicitly employs {}``Occam's razor'' by penalising model
variants with more free parameters. 

\subsection{Fitting}

Each model variant has at least 20 parameters which need to be extracted
by fitting the calculations to the data. Traditional optimizing routines
require a first guess at parameter values. Whilst some prior experience
can be used to estimate the starting values, in a parameter space
of this size it is very difficult to judge whether an optimum found
by an optimizer is local or global. In previous work \cite{Janssen_role_hyp}
a simulated annealing strategy was adopted, however we subsequently
found \cite{jan03} that using a genetic algorithm (GA) in combination
with a traditional optimizer, MINUIT \cite{min95}, offered many advantages.
The GA is able to search a large region of parameter space and very
quickly arrives at reasonable solutions, whereas a minimizer such
as MINUIT, which uses a Davidon-Fletcher-Powell algorithm, is able
to take the solutions from the GA as starting points and find optima,
provided the starting points are not too far from the optimum. This
strategy thus plays to the strengths of both algorithms.

Genetic algorithms are a class of search strategies known as
evolutionary computing. A number of excellent texts on GAs exist
(e.g. \cite{dav91,gol89}), so we will only briefly sketch the strategy
we have employed. The idea is to generate a number of trial solutions
randomly. In this implementation, each solution is an encoding of
trial values of the free parameters as a string of real numbers
$\left\{\lambda_{i}\right\}$, where $\lambda_{i}$ represents the value
of free parameter $i$.

The collection of solutions is referred to as a {}``population''. Each
solution in the population is used to evaluate a function which
determines its {}``fitness''. In our case the fitness function is \[
f\left\{ \lambda_{i}\right\} =\frac{1}{1+\chi^{2}\left\{
\lambda_{i}\right\} },
\]
where $\chi^{2}\left\{ \lambda_{i}\right\} $ is the result of running
the calculation of all the experimental observables and comparing
with the available data.

The population is then {}``evolved'' in a manner analogous to biological
evolution. One or two solutions are selected from the current population,
where solutions with greater fitness function values are selected
preferentially, but not exclusively. When one solution is selected,
it is subjected to a {}``mutation'', where one or more of the free
parameters are altered at random. 

When two solutions are selected, a new individual is created by
{}``crossover'' of the encoded parameters in each solution. In this
implementation, crossover is performed by a number of different
functions, chosen at random, which fall broadly into two
categories. The first is to swap one or more parameters from solution
$\left\{\lambda_{i}\right\}$ to solution $\left\{\mu_{i}\right\}$,
$\lambda_{i}\leftrightarrow\mu_{i}$ two obtain two {}``child''
solutions. The second takes parameters from each solution and forms
one child by assigning it new parameters $\left\{\nu_{i}\right\}$
calculated by a weighted averaged: $\nu_{i} =
\frac{w_{1}\lambda_{i}+w_{2}\mu_{i}}{w_{1}+w_{2}}$.

A new fitness is then evaluated, and if it is better
than the worst current fitness in the population, the new solution
replaces the previous worst. This is often referred to as a steady
state GA. The population therefore gradually migrates to one or more better
points in parameter space, and if the routine is run for long enough,
it will converge on one optimum. Whilst this optimum is not guaranteed
to be global, GA research \cite{gol89} has shown that {}``reasonable''
solutions can be found very efficiently, even when no prior
information has been used to constrain the free parameters. For
instance, in our calculations, the best-of-generation solutions at the
start of a run have $\chi^{2}$ values of order $10^{6}$ but this will
be reduced to of order 10 after only 5000 evaluations of the objective
function. If gradient methods were employed using such a random choice
of initial starting values, it is highly likely they would be trapped
in local minima. Note also that compared to the simulated annealing
strategy used in previous work
\cite{Janssen_role_hyp,Janssen_backgr,Janssen_elec,Janssen_sigma} we
have reduced the number of function evaluations by a factor of 10.
Whereas no one strategy can be optimum for all problems, the GA is
likely to be highly efficient for problems such as the present one
with many parameters and complicated $\chi^{2}$ surfaces. The
best-of-generation solutions at the end of each GA run can be used as
starting points for a MINUIT minimization.

The complete strategy for analysis of each model was as follows:

\begin{enumerate}
\item 100 GA calculations were initiated. Each GA calculation used a population
of 200 and was run for 5000 evaluations of the objective function.
The limits for each parameter were deliberately chosen to define a
large region of parameter space.
\item The best solution from each GA calculation was used as a starting
point for a MINUIT minimization which was run for a further 5000 function
evaluations. At this point very few, if any, of the MINUIT calculations
had achieved convergence and the residual values of $\chi^{2}$ were
too high; a more limited search space was required.
\item The parameters associated with each of the 100 MINUIT solutions were
then examined. The set of solutions whose $\chi^{2}$ values were
within 1.0 of the current best $\chi^{2}$ were selected. A new, tighter
set of limits for each parameter was defined from the range of parameter
values exhibited by the chosen subset of solutions. This typically
reduced parameter ranges by factors of 5-10.
\item 100 new GA calculations were initiated, the only difference from before
being the smaller region of parameter space in which the search was
performed.
\item As before, the best solution from each GA calculation was used to
initiate MINUIT calculations. A few calculations had converged at
this point. It was clear, however, that the calculations had not converged
to the same optimum, so to investigate this, we proceeded to try to
obtain convergence for all the calculations.
\item The MINUIT solutions were then re-inserted back into the GA populations.
Since the inserted solution was likely to be the {}``fittest'' solution
in the population, this would cause the GA to concentrate more on
a particular region of parameter space. Furthermore, the populations
from each of the 100 GA calculations were mixed. This is a variation
on what is known as an island population mixing scheme, and can be
shown to be beneficial in cases where separate GAs are converging
on different optima.
\item A third run of the 100 GA calculations was now performed.
\item A final MINUIT calculation for each of the 100 GA best solutions was
done. At this point, all the calculations had converged.
\end{enumerate}

The result for each model variant was 100 solutions, each of which
represented a converged minimization by MINUIT. As we noted in
\cite{jan03}, we observed that \emph{each solution was unique,} even
although they had roughly similar $\chi^{2}$ values. In other words,
there were \emph{at least} 100 local minima in the $\chi^{2}$ surface.

The fact that each calculation appears to find a different local minimum
shows that the optimization problem may be ill-posed. The reason for
this could be either that the reaction model lacks an ingredient which
is necessary to obtain reproducible minimization results, or that
the experimental data contain systematic uncertainties which render
them inconsistent with each other, or that both the model and the
data are problematic. From this observation, we caution that it is
not possible to extract the magnitudes of coupling constants to high
precision by fitting hadrodynamical models such as ours (and by extension
other similar models) to a relatively small number of photoproduction
data points.

What is observed however, is that the {}``best'' of the calculations
(i.e. those with the lowest $\chi^{2}$ values) tend to cluster around
particular regions in parameter space. Our previous work \cite{jan03}
showed that this still resulted in considerable ambiguities in the
calculations of un-measured observables, and pointed to quantities
such as photon asymmetry as being very useful in reducing the uncertainties
in parameter values. The additional data used in the present work
are the beam polarization data from the SPring-8/LEPS facility \cite{zeg03}
and electroproduction data \cite{moh03,Bebek_74,Bebek_77,Brown_cea},
and it appears they have affected a marked reduction in the regions
of parameter space which give a good fit to the data (see section
\ref{sec:Results-and-discussion}). This is a valuable observation.
It shows that our original work, in agreement with others (e.g. \cite{Mart2}),
was able to predict the measurements which would most significantly
improve the comparison between data and theory. 

\subsection{Model Variant Comparison}

Whilst the best fit for each model was obtained by minimising $\chi^{2}$,
this does not take into account the different number of free parameters
in each model, or how small a region of parameter space corresponds
to a reasonable fit. A full description of the framework we have employed
is given in the appendix, but we sketch the salient points here.

Comparing two model variants ($A$ and $B$, say), we can evaluate
the ratio of the likelihood functions:\begin{equation}
\frac{P\left(D|A\right)}{P\left(D|B\right)}\,,\label{eq:ratios}\end{equation}
where $P\left(D|A\right)$ ($P\left(D|B\right)$ ) is the probability
that the data $D$ would be obtained, assuming that variant $A$ ($B$)
were true. Under some reasonable assumptions (as mentioned previously),
minimising $\chi^{2}$ is equivalent to maximising the likelihood,
but what is actually obtained could be written as $P\left(D|\lambda_{i}^{MP},A\right)$
(where $MP$ stands for {}``most probable''). This expression indicates
that the maximum likelihood depends on the parameters of the model,
$\lambda_{i}$, whereas Eq.\ref{eq:ratios} requires there to be no
dependence on the free parameters of either variant.

Using further approximations, we obtain for variant $A$ the expression
(with a similar one for variant $B$):\begin{equation}
P\left(D|A\right)\propto P\left(D|\lambda^{MP},A\right)\frac{\prod_{i}^{N}\delta\lambda_{i}^{MP}}{\prod_{i}^{N}\left(\lambda_{i}^{max}-\lambda_{i}^{min}\right)}\,,\label{eq:approx_likelihood}\end{equation}
where there are $N$ free parameters, $\lambda_{i}^{max}$ and $\lambda_{i}^{min}$
represent the maximum and minimum limits of search space for parameter
$\lambda_{i}$, and $\delta\lambda_{i}^{MP}$ is the uncertainty in
the most probable value of parameter $\lambda_{i}$. The second factor
in Eq.\ref{eq:approx_likelihood} is often referred to as the {}``Occam
factor'', since it penalizes model variants with more free parameters.
It is most easily interpreted as a product of $N$ factors, each one
a ratio of fitted error bar size to search space size for parameter
$\lambda_{i}$. Model variants whose parameter error bars are relatively
large are thus more favoured by the data since it means that there
is a larger region of parameter space which gives a good fit of the
model to the data. The ratio in equation \ref{eq:ratios} becomes
not just a ratio of maximum likelihoods, but is multiplied by the
ratio of Occam factors.

Coupling constants associated with the Born amplitudes and the cutoff
parameters were free parameters in the fitting procedure, but tended
to migrate to limits imposed by physical considerations such as SU(3)
symmetry. However, for all model variants these parameters were roughly
similar, and so were {}``uninteresting'' from a physical point of
view. We therefore reduced the effective number of free parameters
after the fitting procedure by considering only those parameters which
related are to coupling constants of resonances in the $s$-channel. 

\section{\label{sec:Results-and-discussion}Results and discussion}

\subsection{Fits to the available data}

\begin{figure}
\begin{center}
\includegraphics[%
  scale=0.5]{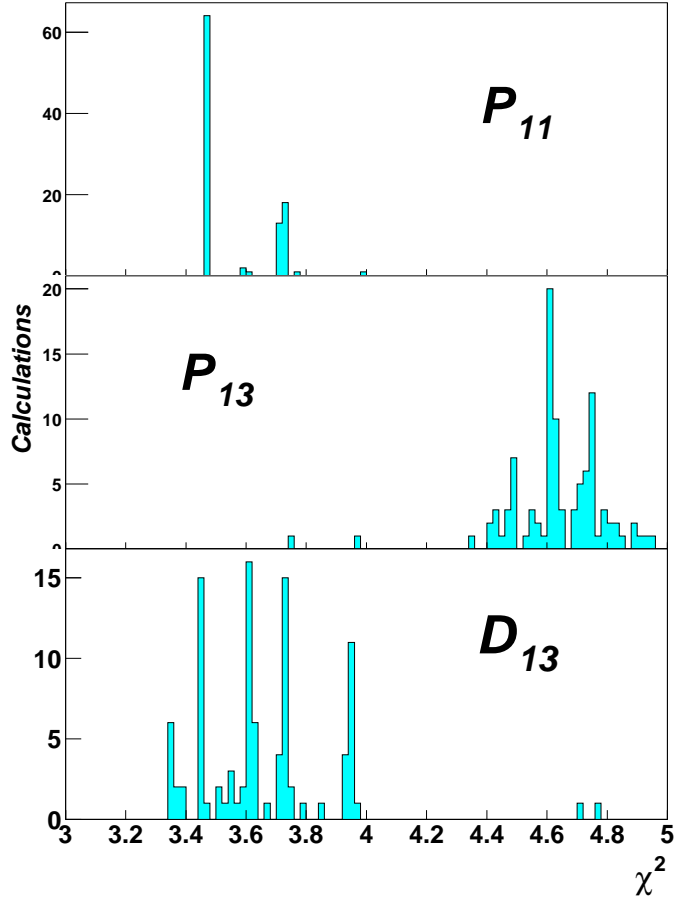}

\caption{\label{cap:Histograms-of-the}Histograms of the number of calculations
resulting in $\chi^{2}$ values}
\end{center}
\end{figure}

The results from the fitting calculations are summarized in table
\ref{cap:Results-from-the}. The factors quoted in the table are described
as follows:

\begin{enumerate}
\item \textbf{$\chi^{2}$}: The value quoted is the lowest $\chi^{2}$ per
data point found for each model variant. 
\item \textbf{Number of free parameters:} This is the number of $s$-channel
resonance coupling constants which were varied in the fitting procedure,
as explained in section \ref{sec:Analysis-Procedure}.
\item \textbf{Number of {}``best'' calculations:} For each model variant,
a histogram of the number of calculations versus $\chi^{2}$ was produced.
Fig. \ref{cap:Histograms-of-the} shows three such histograms representing
the results from the $P_{11}$, $P_{13}$ and $D_{13}$ model variants.
Remembering that \emph{each} calculation is a \emph{converged} MINUIT
result, it can be seen that there is a range of $\chi^{2}$ values
for each variant. Examination of figure \ref{cap:Histograms-of-the}
reveals significant differences in the results according to model
variant. For instance, $P_{11}$ shows a large number of calculations
at one value of $\chi^{2}$ (at the level of histogram binning), whereas
$P_{13}$ shows a much wider range in $\chi^{2}$, but there is one
calculation which has attained a much better $\chi^{2}$ than all
the others, and $D_{13}$ is somewhere in between the two. One can
also see that $D_{13}$ contains calculations with the lowest $\chi^{2}$,
but there are only a handful of them, whilst $P_{11}$ has a large
fraction of its calculations in the lowest $\chi^{2}$ {}``peak''.
The number of best calculations represents the number in the lowest
$\chi^{2}$ {}``peak'' for each model variant. 
\item \textbf{Occam factor:} The Occam factors are also included to show
how the various model variants are penalized for having either a larger
number of free parameters, or which obtain good fits only in a smaller
region of parameter space. They are normalized to the value obtained
for the {}``Core'' variant.
\item \textbf{Ratio of Posterior Probabilities:} The figures quoted are
the result of applying the formulae derived in section \ref{sec:Analysis-Procedure}
and the appendix. As with the Occam factor, they are also normalized
to the value obtained for the {}``Core'' variant.
\end{enumerate}

\begin{table}
\begin{center}
\begin{tabular}{|c|c|c|c|c|c|}
\hline 
Model Variant&
Core&
$S_{11}$&
$P_{11}$&
$P_{13}$&
$D_{13}$\tabularnewline
\hline
\hline 
$\chi^{2}$&
5.14&
4.47&
3.47&
3.75&
3.35\tabularnewline
\hline 
Number of free parameters&
4&
5&
5&
6&
6\tabularnewline
\hline 
Number of {}``good'' calculations&
15&
1&
64&
1&
8\tabularnewline
\hline 
Occam factor&
1.000&
3.278&
5.556&
0.018&
1.167\tabularnewline
\hline 
Ratio of Posterior Probabilities&
1.000&
4.500&
12.571&
0.035&
2.786\tabularnewline
\hline
\end{tabular}

\end{center}
\caption{\label{cap:Results-from-the}Results from the fitting calculations.
The various factors are explained in the text.}
\end{table}

The {}``raw'' $\chi^{2}$ results show that the $D_{13}$ model
variant gives the best fit to data (in agreement with other predictions
\cite{Mart2}), but that the $P_{11}$ and $P_{13}$ variants also
give comparable fits. However, once the Occam factor is taken into
account, the ratio of posterior probabilities show that the $P_{11}$
variant is favoured by a factor of 4.5 over $D_{13}$ and a factor
of 359 over $P_{13}$. \textcolor{black}{It is interesting to note
that this mirrors the {}``intuitive'' feeling we obtained when examining
the distributions of $\chi^{2}$ results shown in figure \ref{cap:Histograms-of-the},
i.e. the model variant with the greater number of low $\chi^{2}$
calculations ($P_{11}$) is favoured over the one with very few ($P_{13}$).
It is also} interesting to note that $S_{11}$ is slightly favoured
over $D_{13}$, even although the raw fit to data is not as good;
this presumably reflects the fact that $D_{13}$ contains more parameters
than $S_{11}$. 

The main points we would like to conclude from this are:

\begin{enumerate}
\item The available data show evidence that, within our model framework,
an additional resonance is required.
\item The quantum numbers of the additional resonance are not possible to
ascertain (given the amount of experimental data), but the present
data set give greatest support to the hypothesis that it is a $P_{11}$.
\end{enumerate}

\subsection{\label{sub:Comparison-with-the}Comparison with the data used in
the fitting procedure}

In order to show how well each model variant can fit the data, we
present a selection of plots which show the comparison. For each variant,
the parameters found by the fitting calculation which produced the
best $\chi^{2}$ value have been used. Our previous work \cite{jan03}
showed calculations using many different fitted parameter sets to
make the point that a great deal of ambiguity remained after fits
to data. However, now that we have included new data points in the
fitting process, namely the photon beam asymmetry and electroproduction
data, the ambiguity in parameters within the same model has been drastically
reduced. We therefore restrict ourselves to plotting one calculation
for each model variant, since the differences within a model variant
are much smaller than the differences amongst the different variants. 

\begin{figure}
\begin{center}
\includegraphics[%
  scale=0.7]{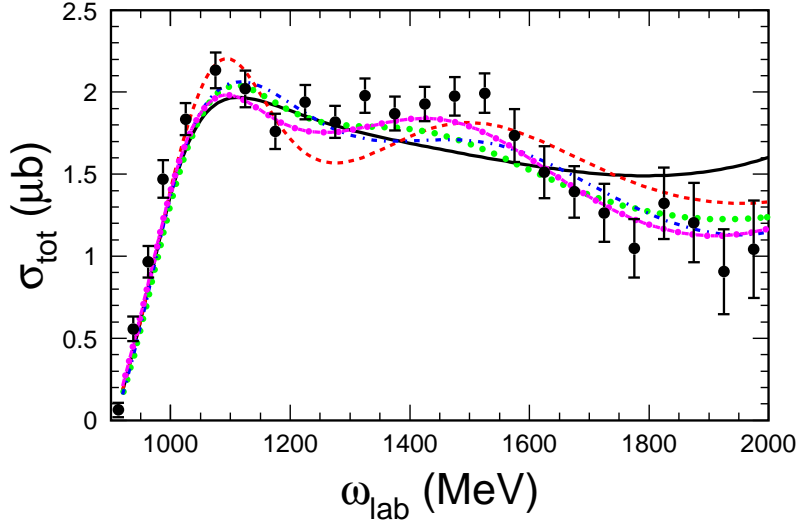}

\caption{\label{cap:Total-cross-section}Total cross-section for the $p(\gamma,K^{+})\Lambda$
reaction. The data points are from \cite{Tran}. The model variant
calculations are: core set (solid line), $S_{11}$ (dashed line),
$P_{11}$ (small circles), $P_{13}$ (dot-dashed line) and $D_{13}$
(solid plus circles).}
\end{center}
\end{figure}

\begin{figure}
\begin{center}
\includegraphics[%
  scale=0.7]{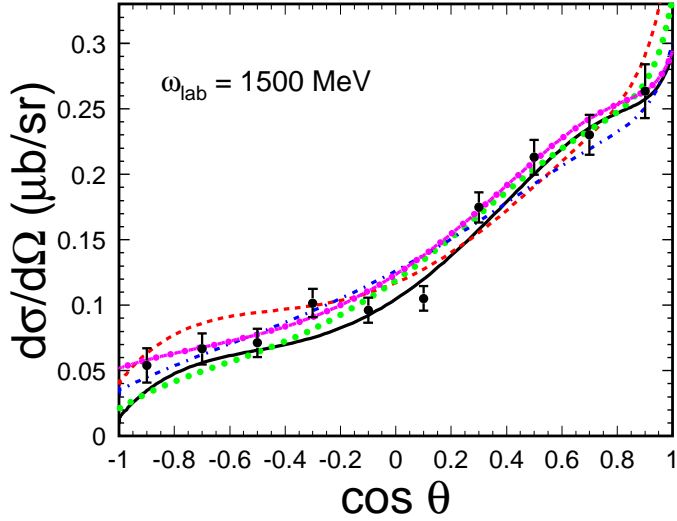}

\caption{\label{cap:Differential}Fit of each model variant to differential
cross-section data. The lines have the same meaning as in figure \ref{cap:Total-cross-section}.}
\end{center}
\end{figure}

\begin{figure}
\begin{center}
\includegraphics[%
  scale=0.7]{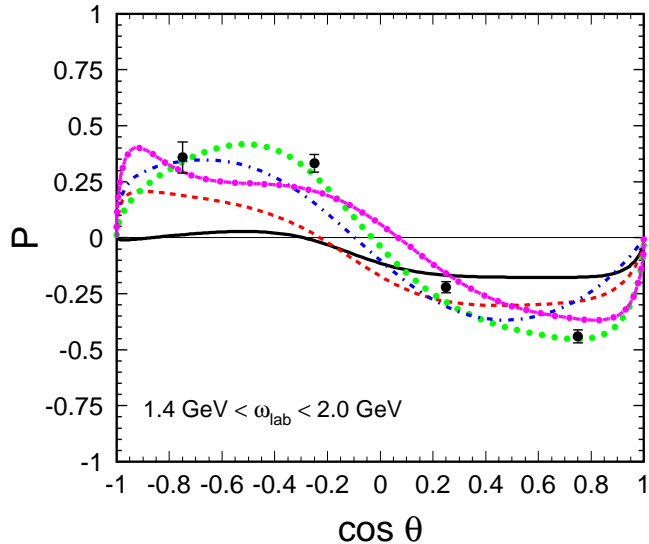}

\caption{\label{cap:Recoil-pol}Fit of each model variant to the recoil polarisation
data. The lines have the same meaning as in figure \ref{cap:Total-cross-section}.}
\end{center}
\end{figure}

Figure \ref{cap:Total-cross-section} shows the comparison of all
the model variants to the total $p(\gamma,K^{+})\Lambda$ cross-section
data, whilst figures \ref{cap:Differential} and \ref{cap:Recoil-pol}
show typical differential cross-section and recoil polarization fits
respectively. At this point we note that the four model variants which
include an additional resonance are able to reproduce a {}``resonant
bump'' in the total cross-section, whilst the core variant cannot.

The recoil polarizaton in figure \ref{cap:Recoil-pol} again shows
that an additional resonance is required in order to reproduce the
data especially at backward angles. Whilst a {}``bump'' in the total
cross-section was reproduced by Saghai \cite{Saghai_hyp_res} by altering
an ingredient of the background processes, it is unlikely that the
tuning of the background could result in a substantial non-zero recoil
polarization as seen in the data. This observation may be a further
indication that an additional resonance is required in this energy
range.

\begin{figure}
\begin{center}
\includegraphics[%
  scale=0.7]{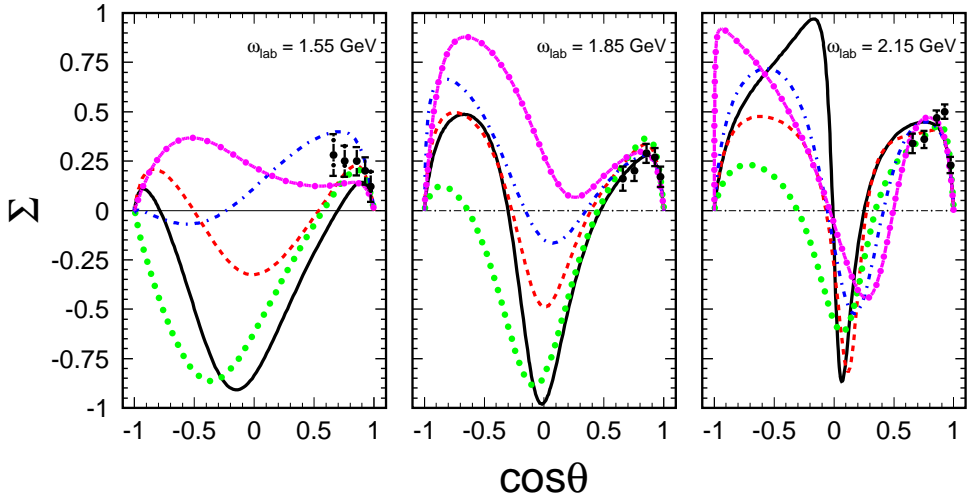}

\caption{\label{cap:Photon-asymmetry}Fit of each model variant to photon
beam asymmetry data. The data are from \cite{zeg03}, and the lines
have the same meaning as in figure \ref{cap:Total-cross-section}.}
\end{center}
\end{figure}

Figure \ref{cap:Photon-asymmetry} shows the calculations compared
to the recent SPring-8 photon asymmetry \cite{zeg03}. It is clear
that these data have greatly influenced the overall fit, since they
tie down all the calculations at forward angles. Indeed, we predicted
that this would be the case \cite{jan03}. However, especially at
the lower energy setting (1.55 GeV), it is still the case that there
is a large difference in the model calculations over most of the angular
range. Measurements over an extended range of angle would therefore
be very useful to constrain further the fitting procedure. At 1.55
GeV, even an asymmetry averaged over the entire angular range would
discriminate between the $P_{11}$ and $D_{13}$ calculations, since
the predictions are negative and positive respectively.

\begin{figure}
\begin{center}
\includegraphics[%
  scale=0.6]{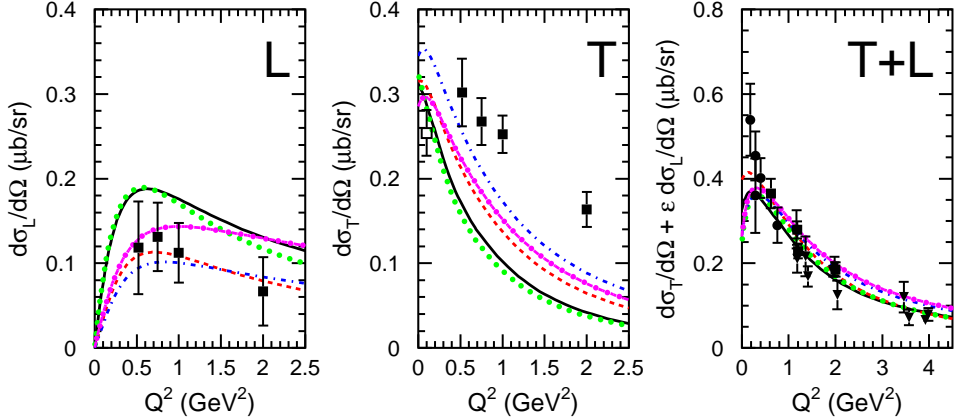}

\caption{\label{cap:Electroproduction}Fits of model variants to electroproduction
data. The separated longitudinal (L) and transverse (T) differential
cross section filled square data are from \cite{moh03} at $W=1.84$GeV
and $\theta_{qK}^{\star}=0^{\circ}$, and the $Q^{2}=0$ (photoproduction
point) from \cite{Tran} is shown as a comparison. The combined (T+L)
differential cross-section data are from \cite{Bebek_74,Bebek_77,Brown_cea}
at $W=2.15$GeV and $\theta_{qK}^{\star}=8^{\circ}$. The lines have
the same meaning as in figure \ref{cap:Total-cross-section}.}
\end{center}
\end{figure}

Comparisons with electroproduction data are presented in figure \ref{cap:Electroproduction}.
Longitudinal, transverse and combined differential cross-sections
are shown. It is clear that agreement with the transverse cross section
is the least good of all the comparisons with data, and is probably
responsible for the residual high values of $\chi^{2}$ found in the
fitting process. However, we note that there is virtually no difference
amongst the model variants in the combined cross-sections, and only
a small scale and form difference in the separated cross-sections.
In other words, these electroproduction observables are relatively
insensitive to details of the resonances in the reaction, and mirror
our previous work \cite{Janssen_elec} where we concluded that they
are highly influenced by {}``background'' processes. 

\begin{figure}
\begin{center}
\includegraphics[%
  scale=0.7]{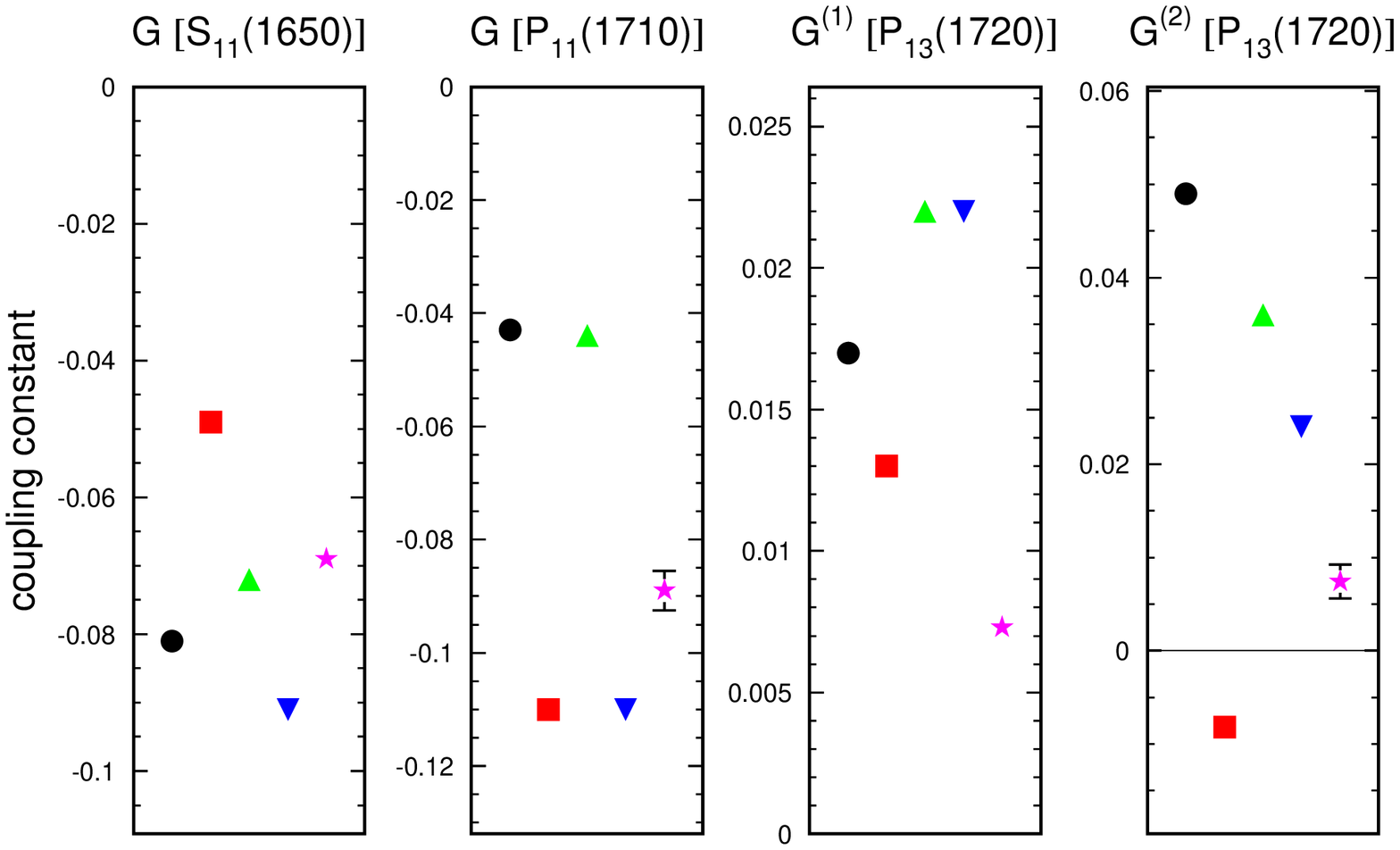}

\caption{\label{cap:Coupling-constants-plot}Depiction of the values of the
main coupling constant found for each model variant. The meaning of
the symbols are: circle - core set, square - $S_{11}$, upright triangle
- $P_{11}$, downwards triangle - $P_{13}$ and star - $D_{13}$. }
\end{center}
\end{figure}

For completeness, and to summarize the results of our analysis, we
show in table \ref{cap:Coupling-constant-table} the values of the
resonance coupling constants which correspond to the best fit calculation
for each model variant. In addition, we show graphically the range
and correlation of the coupling constant values in figure \ref{cap:Coupling-constants-plot},
for those coupling constants which are common to all models%
\footnote{Note that in figure 4 of reference \cite{jan03}, there was an error
in the normalisation of the $P_{13}(1720)$ and $D_{13}(1895)$ coupling
constants. The values used in the calculations of that article were
correct. %
}. Error bars (where shown) indicate the spread (standard deviation)
of the values for the {}``best'' calculations, defined in table
\ref{cap:Results-from-the}, otherwise the spread is less than the
size of the symbol. This plot indicates that the values extracted
from the process are indeed highly correlated with the choice of additional
resonance, as there are no discernable similarities in values and
there is also a sizeable spread.

\begin{table}
\begin{center}
\begin{sideways}
\begin{tabular}{|c||c|c|c|c|c|}
\hline 
&
\multicolumn{5}{c|}{Model Variant}\tabularnewline
\hline 
Coupling Constant&
Core&
$S_{11}$&
$P_{11}$&
$P_{13}$&
$D_{13}$\tabularnewline
\hline
\hline 
$G[S_{11}(1650)]$&
$-8.1\times10^{-2}\pm3.6\times10^{-5}$&
$-4.9\times10^{-2}$&
$-7.2\times10^{-2}\pm2.1\times10^{-4}$&
$-9.1\times10^{-2}$&
$-6.9\times10^{-2}\pm2.2\times10^{-4}$\tabularnewline
\hline 
$G[P_{11}(1710)]$&
$-4.3\times10^{-2}\pm9.4\times10^{-5}$&
$-1.1\times10^{-1}$&
$-4.4\times10^{-2}\pm1.4\times10^{-3}$&
$-1.1\times10^{-1}$&
$-8.9\times10^{-2}\pm3.5\times10^{-3}$\tabularnewline
\hline 
$G^{(1)}[P_{13}(1720)]$&
$1.7\times10^{-2}\pm1.2\times10^{-5}$&
$1.3\times10^{-2}$&
$2.2\times10^{-2}\pm1.6\times10^{-4}$&
$2.2\times10^{-2}$&
$7.3\times10^{-3}\pm2.8\times10^{-4}$\tabularnewline
\hline 
$G^{(2)}[P_{13}(1720)]$&
$4.9\times10^{-2}\pm6.0\times10^{-5}$&
$-8.2\times10^{-3}$&
$3.6\times10^{-2}\pm6.9\times10^{-4}$&
$2.4\times10^{-2}$&
$7.4\times10^{-3}\pm1.8\times10^{-3}$\tabularnewline
\hline 
$G[S_{11}(1895)]$&
-&
$8.7\times10^{-2}$&
-&
-&
-\tabularnewline
\hline 
$G[P_{11}(1895)]$&
-&
-&
$-3.8\times10^{-1}\pm5.5\times10^{-4}$&
-&
-\tabularnewline
\hline 
$G^{(1)}[P_{13}(1895)]$&
-&
-&
-&
$2.1\times10^{-2}$&
-\tabularnewline
\hline 
$G^{(2)}[P_{13}(1895)]$&
-&
-&
-&
$2.1\times10^{-2}$&
-\tabularnewline
\hline 
$G^{(1)}[D_{13}(1895)]$&
-&
-&
-&
-&
$-2.2\times10^{-1}\pm5.7\times10^{-4}$\tabularnewline
\hline 
$G^{(2)}[D_{13}(1895)]$&
-&
-&
-&
-&
$-2.5\times10^{-2}\pm8.0\times10^{-4}$\tabularnewline
\hline
\end{tabular}
\end{sideways}

\end{center}

\caption{\label{cap:Coupling-constant-table}The values of the coupling constants
extracted from the fitting procedures.}
\end{table}

\subsection{Comparisons with other data}

\begin{figure}
\begin{center}
\includegraphics[%
  scale=0.6]{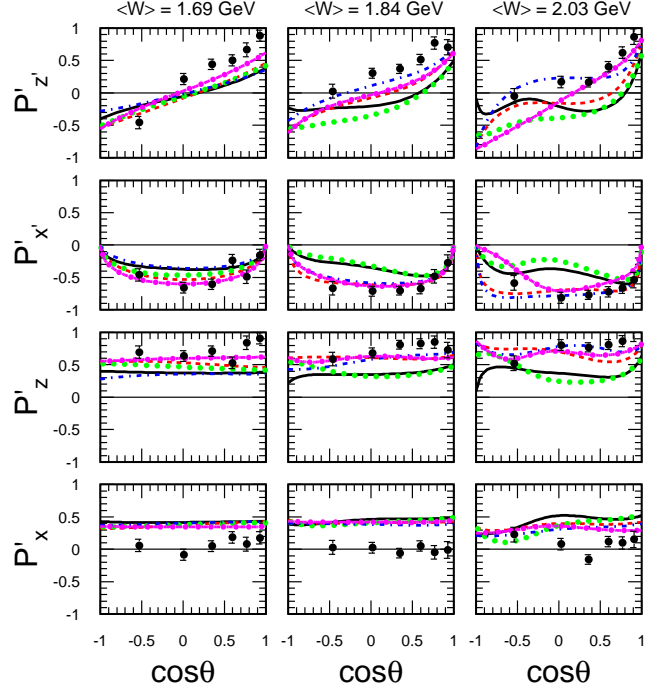}

\caption{\label{cap:Dan-Carman-electroproduction}Comparison with the double
polarization electroproduction data of ref. \cite{car03}. The line
conventions are the same as Fig. \ref{cap:Total-cross-section}.}
\end{center}
\end{figure}

A recent experiment using the CLAS detector at Jefferson Lab \cite{car03}
has measured the transferred polarization in the exclusive $p(\vec{e},e'K^{+})\vec{\Lambda}$
reaction. Figure \ref{cap:Dan-Carman-electroproduction} shows the
comparison of the model variants with this data set. We have not used
this data in the fitting routine because the calculation of each point
would require an averaging over the phase-space of each data bin,
thereby significantly increasing the computational time. Instead we
show the predictions for each model variant, based on the extracted
free parameters described in \ref{sub:Comparison-with-the}. All the
calculations show moderate agreement with the data, with the exception
of the $P_{x}'$ observable which was measured to be consistent with
zero, where none of the model variants appear to be able to reproduce
this behaviour. It was suggested in \cite{car03} that this may be
due to the production of $s\bar{s}$ quark pairs with anti-aligned
spins. This is beyond the scope of the present hadrodynamical model.
However, in common with the separated longitudinal and transverse
differential cross-sections, there is relatively little sensitivity
to details of which resonances are included in the reaction calculations,
which indicates that this reaction is not best one to search for missing
resonances. 

\subsection{Predictions for other observables}

We have already seen that a more comprehensive measurement of the
photon beam asymmetry would be highly sensitive to the choice of resonance
in the reaction. With the prospect of double polarization photoproduction
experiments \cite{Klein}, we present some predictions, based on the
models described above and using the parameters found in the fitting
procedures. The authors of \cite{Klein} have proposed to measure
beam-target, target-recoil and beam-recoil polarizations for both
$p(\gamma,K^{+})\Lambda$ and $p(\gamma,K^{+})\Sigma$ reactions at
a photon beam energy of 1.5 GeV.

\begin{figure}
\begin{center}
\includegraphics[%
  scale=0.6]{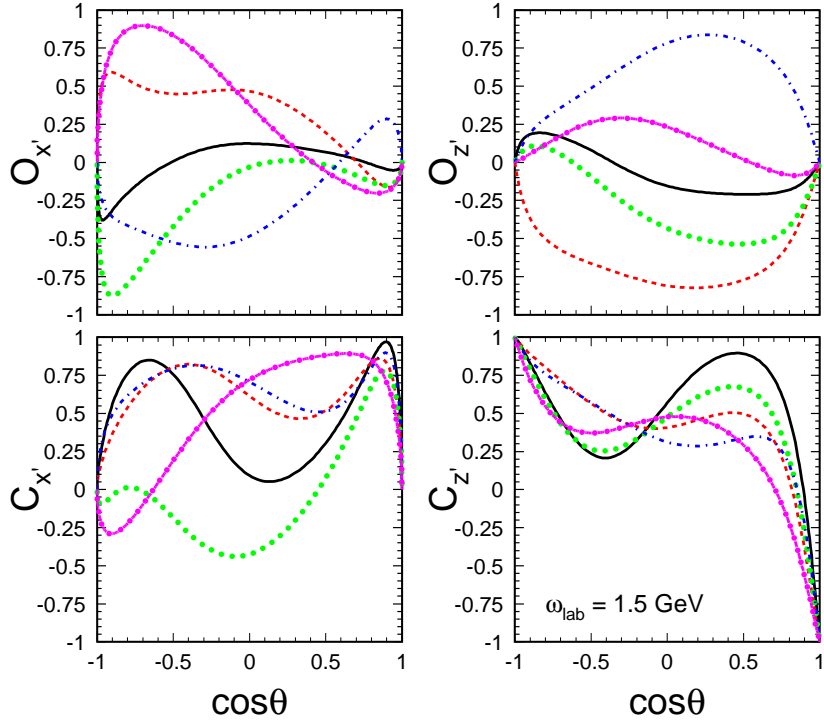}

\caption{\label{cap:Beam-recoil-polarisation-prediction}Beam-recoil polarization
prediction for linearly (top panels) and circularly (bottom panels)
polarized photons.}
\end{center}
\end{figure}

\begin{figure}
\begin{center}
\includegraphics[%
  scale=0.6]{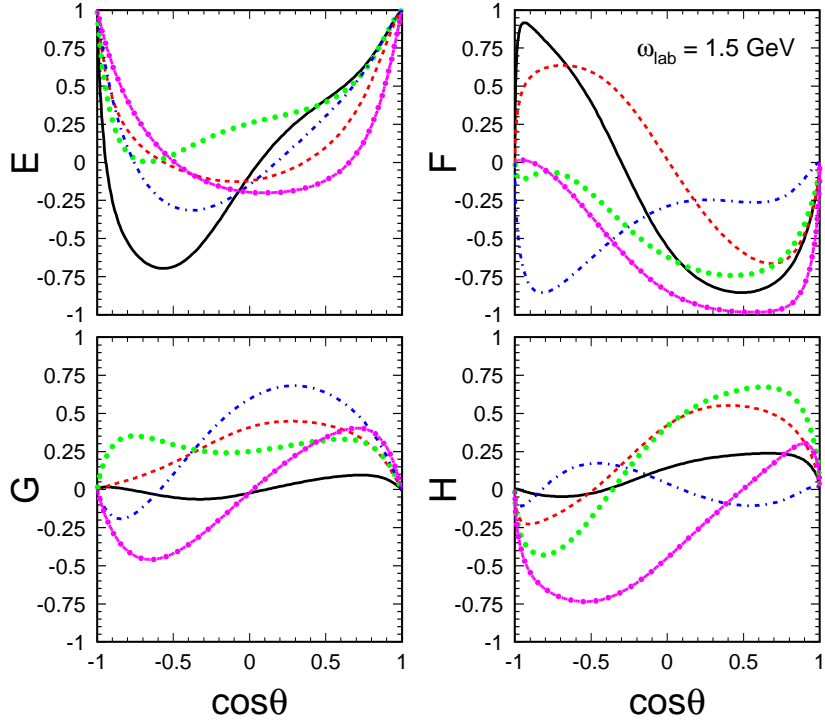}

\caption{\label{cap:Beam-target-polarisation-prediction}Beam-target polarization
prediction for circularly (top panels) and linearly (bottom panels)
polarized photons.}
\end{center}
\end{figure}

Figures \ref{cap:Beam-recoil-polarisation-prediction} and \ref{cap:Beam-target-polarisation-prediction}
show the results of the calculations using each of our models of beam-recoil
and beam-target polarization for the $p(\gamma,K^{+})\Lambda$ reaction
only. Beam-target, target-recoil and beam-recoil polarizations are
not independent, although in practice the measurement of all three
would be desirable to limit systematic effects. All the observables
show a high degree of sensitivity to the choice of resonance. One
could therefore imagine that the measurement of a few selected points
would be highly valuable in determining the dynamics of the reaction.

\section{\label{sec:Conclusion}Conclusion}

A framework for analysing $p(\gamma,K^{+})\Lambda$ reaction data
sets has been presented. Using the foundation of a tree-level hadrodynamical
model, we have shown that the extraction of coupling constants by
fitting to data is not trivial. The strong suggestion from our work
is that fitting procedures should be carried out several times in
order to establish how reliable the extracted numbers are. Carrying
out the fitting procedure using a hybrid of a genetic algorithm and
a traditional minimizer has been shown to play to the strengths of
both strategies, and carrying out many calculations in parallel allows
a thorough investigation of the character of the search space. Any
conclusions from such procedures must take into account the uncertainty
resulting from calculations which converge on different optima in
parameter space.

From the analysis shown here, we have reasonable evidence that an
extra resonance is required to explain the $p(\gamma,K^{+})\Lambda$
reaction data within our model. The study shows that the data
currently favour the presence of an additional $P_{11}$ resonance of
mass 1895 MeV, but that $S_{11}$ and $D_{13}$ resonances are also
weakly supported and cannot therefore be ruled out. We have not
studied how combinations of additional resonances might improve the
description. A firm conclusion awaits further data and a full-blown
theoretical coupled-channel analysis to pin down the remaining
uncertainty. There is a strong possibility that a GA may be of great
value in the minimization required in a coupled-channel approach.

We have shown that whilst the different model variants do a reasonable
job of describing the $p(e,e'K^{+})\Lambda$ data, they are relatively
insensitive to the details of the contributing resonances. The polarization
observables in photoproduction reactions remain the best candidates
for investigating the nature of any additional resonances. 

We are aware that further comprehensive data sets have been published
\cite{gla03,mcn04}. We look forward to being able to use this
to improve the quality of the information extracted in our approach.
In addition, the inclusion of radiative kaon capture data
\cite{pha01} may provide further constraints on the values of
resonance couplings. However, we note that in this reaction, which is
crossed-symmetric to the $p(\gamma,K^{+})\Lambda$ studied above,
coupled channel effects have been shown to be important
\cite{sie95,lee98}.

This work was supported by the UK's Engineering and Physical Sciences
Research Council, and the Fund for Scientific Research - Flanders. 

\section*{Appendix: Model comparison}

We compare the relative success of the different model variants by
evaluating the ratios of the maximum posterior probabilities \cite{siv96}.
To compare variants $A$ and $B$ the ratio, after application of
Bayes Theorem, becomes

\[
\frac{P\left(A|D\right)}{P\left(B|D\right)}=\frac{P\left(D|A\right)P\left(A\right)}{P\left(D|B\right)P\left(B\right)}\,,\]
where $P\left(A|D\right)$ is the maximum posterior probability for
variant $A$, $P\left(D|A\right)$ is the probability that the data
would be obtained, assuming variant $A$ to be true, $P\left(A\right)$
is the \emph{a priori} probability that variant $A$ is correct and
similarly for variant $B$. With no prior prejudice as to which variant
is correct, we obtain the ratio of likelihoods:\[
\frac{P\left(D|A\right)}{P\left(D|B\right)}\,.\]

The likelihood $P\left(D|A\right)$ is an integral over the joint
likelihood $P\left(D,\left\{ \lambda_{i}\right\} |A\right)$, where
$\left\{ \lambda_{i}\right\} $ represents a set of free parameters:\begin{equation}
\begin{array}{ccl}
P\left(D|A\right) & = & \int P\left(D,\lambda_{i}|A\right)d\lambda_{i}\\
 & = & \int P\left(D|\lambda_{i},A\right)P\left(\lambda_{i}|A\right)d\lambda_{i}\end{array}\,,\label{eq:likelihood}\end{equation}
where we have dropped the curly brackets for convenience. Model variants
may have different numbers of parameters, and the multi-dimensional
integral in equation \ref{eq:likelihood} is the formal means of handling
the problem. The function $P\left(\lambda_{i}|A\right)$ is the prior
probability that the parameters take on specific values. With no prior
prejudice, we assume that each parameter $\lambda_{i}$ lies in the
range $\lambda_{i}^{min}\leq\lambda_{i}\leq\lambda_{i}^{max}$, and
we can write the prior as the reciprocal of the volume of a hypercube
in parameter search space (N-dimensional for N free parameters):\[
P\left(\lambda_{i}|A\right)=\frac{1}{\prod_{i}^{N}\left(\lambda_{i}^{max}-\lambda_{i}^{min}\right)}\,.\]
To evaluate the volume of the hypercube in parameter space for each
variant, the limits defined after the first GA plus MINUIT calculations
were used, and in general they were different for each variant.

The factor $P\left(D|\lambda_{i},A\right)$ in the integrand of equation
\ref{eq:likelihood} is not trivial to evaluate. If the errors in
the fitted parameters, $\delta\lambda_{i}^{MP}$ corresponding to
the maximum likelihood are multivariate Gaussian, it can be shown
that\begin{equation}
P\left(D|\lambda_{i},A\right)\propto P\left(D|\lambda_{i}^{MP},A\right)Det^{-\frac{1}{2}}\left(\frac{\mathbf{H}}{2\pi}\right)\,,\label{eq:likelihood2}\end{equation}
where $H=-\bigtriangledown^{2}\ln P\left(\lambda|D,A\right)$ is the
Hessian matrix, and is the inverse of the error matrix obtained in
the fitting process which yields $\lambda^{MP}$. The explanations
for this are given in standard texts on Bayesian data analysis \cite{siv96}.

Equation \ref{eq:likelihood2} can thus be written as\begin{equation}
P\left(D|A\right)\propto P\left(D|\lambda_{i}^{MP},A\right)\frac{Det^{-\frac{1}{2}}\left(\frac{\mathbf{H}}{2\pi}\right)}{\prod_{i}^{N}\left(\lambda_{i}^{max}-\lambda_{i}^{min}\right)}\,,\label{eq:likelihood3}\end{equation}
where the second factor is the Occam factor. We make a further simplification
in our case. As mentioned in the main text, some of the free parameters,
such as those respresenting coupling constants associated with Born
amplitudes and cutoff parameters tended to migrate to limits. Apart
from being physically {}``uninteresting'', this also caused the
determinant of the error matrix to be very sensitive to details of
these fitted parameters, since the errors on each were very small.
The physically interesting parameters are the ones which relate to
resonances in the $s$-channel. 

Rather than re-doing the whole minimization procedure with fixed Born
amplitudes and cutoffs (which we deemed unnecessary due to the likely
inaccuracy of the results from fitting to a limited data set), we
took the errors on the most probable parameter values from the MINUIT
calculations. MINUIT returns values which take into account the effect
of correlations amongst parameters, so we applied the approximation\[
Det^{-\frac{1}{2}}\left(\frac{\mathbf{H}}{2\pi}\right)=\prod_{i}^{M}\delta\lambda_{i}^{MP}\,,\]
where $M$ is the reduced number of parameters.

The Occam factor for each variant was therefore reduced to \[
\frac{\prod_{i}^{M}\delta\lambda_{i}^{MP}}{\prod_{i}^{M}\left(\lambda_{i}^{max}-\lambda_{i}^{min}\right)}\,.\]
The first factor in equation \ref{eq:likelihood3} can be evaluated
up to a scale factor by \[
P\left(D|\lambda^{MP},A\right)\propto\exp\left(-\frac{\chi^{2}}{2}\right)\,.\]
Hence, by taking a ratio of the total likelihoods for two variants,
all common factors will divide out, leaving a figure which represents
the relative probability of two variants being correct.

\end{document}